# Extremely high mobility over 5000 cm$^2$/Vs obtained from MoS$_2$ nanosheet transistor with NiO$_x$ Schottky gate

Hee Sung Lee[1], Seung Su Baik[1,3], Sung-Wook Min[1], Pyo Jin Jeon[1], Jin Sung Kim[1], Kyujin Choi[1], Sunmin Ryu[2], Hyoung Joon Choi[1,3], Jae Hoon Kim[1] and Seongil Im[1*]

[1] Institute of Physics and Applied Physics, Yonsei University, Seoul 120-749, Korea

[2] Department of Applied Chemistry, Kyung Hee University, Yongin, Gyeonggi 446-701, Korea

[3] Center for Computational Studies of Advanced Electronic Material Properties, Yonsei University, Seoul, 120-749, Korea.

*Correspondence to: semicon@yonsei.ac.kr

**Abstract**:

Molybdenum disulfide ($MoS_2$) nanosheet, one of two dimensional (2D) semiconductors, has recently been regarded as a promising material to break through the limit of present semiconductors including graphene. However, its potential in carrier mobility has still been depreciated since the field-effect mobilities have only been measured from metal-insulator-semiconductor field effect transistors (MISFETs), where the transport behavior of conducting carriers located at the insulator/$MoS_2$ interface is unavoidably interfered by the interface traps and gate voltage. Here, we for the first time report $MoS_2$-based metal semiconductor field-effect transistors (MESFETs) with $NiO_x$ Schottky electrode, where the maximum mobilities or carrier transport behavior of the Schottky devices may hardly be interfered by on-state gate field. Our MESFETs with single-, double-, and triple-layered $MoS_2$ respectively demonstrate high mobilities of 6000, 3500, and 2800 $cm^2$/Vs at a certain low threshold voltage of -1 ~ -2 V. The thickness-dependent mobility difference in MESFETs was theoretically explained with electron scattering reduction mechanisms.

Molybdenum disulfide ($MoS_2$) layers recently appeared as an alternative nanosheet material that may overcome a drawback of graphene, whose energy band gap ($E_g <$ ~100 meV) is too small to be used as a transistor channel *(1)*. Bulk $MoS_2$ is known to have an indirect bandgap of ~1.2 eV, but a few angstrom-thin single-layered $MoS_2$ has been reported to exhibit a direct bandgap of 1.8 eV *(2, 3)*. According to the literature, a few-layered $MoS_2$ has displayed high field-effect mobilities in top- and bottom-gate metal-insulator-semiconductor field-effect transistors (MISFETs) *(4)*. Highest mobility has been mostly achieved from single layer $MoS_2$ MISFETs with a top-gate high-*k* $HfO_x$ dielectric layer, but it is in a wide range from 100 to 600 $cm^2/V$ s along with different threshold voltages because the mobility should be seriously dependent upon the dielectric/$MoS_2$ interface states in a MISFET and influenced by the gate voltage as well *(4)*. $MoS_2$ thickness-dependent mobility has also been reported, but needs more clarification *(5)*. So, exact information on the potential mobility of $MoS_2$ nanosheets is still in lack. Here, we introduce $MoS_2$-based metal semiconductor field-effect transistors (MESFETs) with $NiO_x$ Schottky electrode, where the maximum mobilities of the Schottky devices might not be much interfered by the gate/channel interface and gate bias unlike the case of MISFETs *(6, 7)*. Our MESFETs with single- (1L), double- (2L), triple- (3L), ten- (10L), and sixty-layered (60L) $MoS_2$ respectively demonstrate mobilities of ~6000, 3500, 2800, 600, and ~200 $cm^2/Vs$, which are an order of magnitude higher than those of reported MISFETs. Our devices particularly exhibit a certain low operation voltage, following the unique characteristics of MESFET.

Figure 1A displays the top view optical microscopy (OM) image of the MESFET with 1L $MoS_2$ channel contacting 150 nm-thick thermally-evaporated $NiO_x$ gate (G) while the inset shows the scanning electron microscopy (SEM) of the same device on $SiO_2$/$p^+$-Si substrate. Similar OM image of another MESFET but with 3L $MoS_2$ on glass substrate is shown in Figure 1B, where it is clearly noted that our $NiO_x$ gate is transparent enough to show $MoS_2$ channel underneath. $NiO_x$ is known to have quite a deep work function of more than 5.1~5.2 eV as a Ni-rich semitransparent conducting oxide. For a source/drain (S/D) ohmic contact, we utilized an Au/Ti bilayer, so that Ti might contact our $MoS_2$ nanosheet, as shown in the schematic three-dimensional (3-D) view of our MESFET device (Fig. 1E). The channel thickness of our 1L $MoS_2$ on $SiO_2$/$p^+$-Si was confirmed by Raman spectroscopy in fig. S1. The layer numbers of our thick $MoS_2$ sheets were estimated to be 10 and 60, respectively as the thicknesses were measured to be ~7 and ~40 nm by atomic force microscope (AFM) scan (Fig. 1, C and D, where each inset displays topographical 3-D images. See fig. S2 for OM images of thick $MoS_2$ nanosheets).

In order to achieve any typical operation of MESFET, Schottky barrier should be clearly formed between the gate and channel semiconductor nanosheet, so that the G-to-S or G-to-D Schottky diode operates as a rectifier. Secondly, the MESFET should turn on before the Schottky diode starts to operate, otherwise drain current ($I_D$) would be consumed by or leaked to the gate (joining the G-to-S leakage current, $I_{GS}$) (fig. S3). According to the current-voltage (I-V) characteristics from the 1-, 10-, and 60-layered $MoS_2/NiO_x$ diode junction, the 40 nm-thick $MoS_2$ (with 60L) was not able to properly form any good Schottky junction with $NiO_x$ showing very small ON/OFF ratio of less than 10 (Fig. 2A). On the other hand, 1L and 10L $MoS_2$ sheets appear to respectively form good Schottky barriers with respect to $NiO_x$. (so the respective Schottky diodes show 1.32 and 2.49 as their ideality factors ($\eta$).) It is also recognized that the reverse leakage current becomes larger with the sheet thickness, indicating that some lowering of physical barrier takes place with a thick $MoS_2$. The Schottky barrier lowering is certainly related to the thickness-induced band-gap reduction in $MoS_2$; the conduction band edge of $MoS_2$ might decrease with the thickness increase as its bandgap does decrease *(3)* (See schematic illustrations and Schottky equations in fig. S4 and fig. S5.). Corresponding to the results (Fig. 2A), the drain current-gate voltage ($I_D$-$V_G$) transfer characteristics are respectively achieved from the three $MoS_2$ nanosheet-based MESFETs as shown in Figure 2B, where MESFET operations are nicely observed from 1L and 10L $MoS_2$ devices as induced by the $V_G$-dependent charge-depletion-modulation (,which is not just to the thickness direction but also to in-plane direction) *(8)*. The 1L $MoS_2$ MESFET shows an ON/OFF ratio higher than ~$10^6$ due to a low OFF-state current of ~1 pA, along with an excellently small subthreshold swing (*SS*) of ~62 mV/dec and a stabilized ON-current $I_D$ of 1.1 µA under 1 V of $V_D$. The small SS which approaches to room temperature ideal value (60 mV/dec) is the typical character of monolayer nanosheet FET. According to the 1L $MoS_2$ transfer curves and the inset $I_D$-$V_D$ output curve, $I_D$ linearly increases with $V_D$, but the transfer characteristics such as SS and ON/OFF ratio are maintained with little regard to $V_D$ increase (Fig. 2C). Figure 2D displays the transfer curves of 10L $MoS_2$-based MESFET, according to which its ON-current $I_D$ (1.6 µA) appears slightly higher than that of 1L channel MESFET but with degraded ON/OFF ratio, OFF current, and SS properties respectively as ~$10^5$ and ~30 pA, and 120 mV/dec. (We also show output curves in the inset). On the one hand, the 60L $MoS_2$ MESFET displays a resistor type behavior, where its $I_D$ appears almost independent of $V_G$ although somewhat higher ON-current of ~2.7 µA was observed due to the thick MoS$_2$ channel (Fig. 2B).

One of the most important information from above Figs. 2, A to B would be the mobility ($\mu$) values of the three FETs. We adopted more general method to estimate $\mu$ here (fig. S6). Now then, the channel mobility equation of MESFET can be expressed with transconductance, $g_m$ as follows *(9, 10)*,

$$g_m = \frac{\partial I_D}{\partial V_G} = \frac{qN_d\mu tW}{L}$$

So, the maximum channel mobility can be calculated as follows,

$$\mu = \frac{Lg_{Max}}{qN_d tW},$$

where $N_d$ is carrier density as number per cm$^3$, $q$ is an electronic charge, $t$, $W$ and $L$ are the thickness, width and length of the channel, respectively. Whether it is linear or saturation regime mobility, the most important information to determine the mobility is the electron carrier density, $n$ or unintentional doping concentration, $N_d$, according to above equations. $N_d$ is reported to be in a range from ~ $5 \times 10^{15}$ to $5 \times 10^{16}$ cm$^{-3}$ in general MoS$_2$ bulk crystals *(11, 12)*, however, we here estimated the values of our own MoS$_2$ sheets by using three methods: the Padovani-Stratton parameter method *(13, 14)* on 10L MoS$_2$ on SiO$_2$/p$^+$-Si (fig. S7 and Table S1), reflection terahertz time-domain spectroscopy (RTHz-TDS) on our original free-standing 60 µm-thick bulk MoS$_2$ specimen (fig. S8), and finally conventional Hall measurements for 20 nm-thick MoS$_2$ on glass (fig. S9); we thus obtained respective *n* values of $1.42 \times 10^{16}$, $1.1 \times 10^{16}$, and $2.22 \times 10^{16}$ cm$^{-3}$. (These electron density values are $1 \times 10^{10}$ ~ $5 \times 10^{10}$ cm$^{-2}$ in area density, which is about two orders of magnitude lower than that of gate charging state in MISFET.) *(15)* Particularly, large size 20 nm-thick (~30L) Hall measurement samples successfully provide *n* of ~$2.22 \times 10^{16}$ cm$^{-3}$ along with a mobility of ~130 cm$^2$/Vs through several times of repetitive measurements, as fabricated on glass. The Hall mobility value is quite comparable to those reported for thick bulk MoS$_2$ in literature *(16, 17)*. If we assume that the present three nanosheets have the same *n* and the dielectric constant $\varepsilon_r$ (although the $\varepsilon_r$ increases with the MoS$_2$ thickness according to previous reports) *(18)*, the mobility $\mu$ is worked out; calculated linear mobility values are then surprisingly high, at least to be 6400, 450~600, and 191 cm$^2$/Vs for 1L, 10L, and 60LMoS$_2$ MESFETs, respectively with a fixed (maximum) n value of $2.22 \times 10^{16}$ cm$^{-3}$. The mobility of thick bulk MoS$_2$ becomes similar; the Hall mobility of 20 nm-thick MoS$_2$ was ~130 cm$^2$/Vs, while is comparable to that of 60L-thick MoS$_2$ in MESFET (191 cm$^2$/Vs) (Fig. 2B). In fact, the mobility values appeared much higher than those of previously reported MISFETs by an order of magnitude, since 1L MoS$_2$ MISFETs have generally displayed much less than 1000 cm$^2$/Vs (in a reported range of 100 ~ 600 cm$^2$/Vs) while multilayer MoS$_2$ showed even much smaller values than those of single layer.

To further confirm and validate above low n and high μ, we also fabricated similar types of 2L and 3L MoS$_2$ MESFETs on glass substrate by using the same exfoliation method, since we wanted to remove any possible parasitic effects by SiO$_2$/p$^+$-Si substrate *(15)*. According to the respective transfer curves of 2L and 3L MoS$_2$ MESFETs on glass (Fig. 2, E and F), their $I_D$-$V_G$ behavior is almost the same as that 1L and 10L MoS$_2$ MESFETs on SiO$_2$/p$^+$-Si (Fig. 2, C and D), except the threshold voltages which are more negative on glass by -1 V. The insets show the transconductance plots to estimate the maximum field-effect mobility of our glass-substrate MESFETs, that turned out to be 3490 and 2815 cm$^2$/Vs at $V_D$ = 1 V for 2L and 3L MESFETs, respectively when we used $2.22 \times 10^{16}$ cm$^{-3}$ for electron density. Hence, it is regarded that the NiO$_x$ Schottky-driven MoS$_2$ MESFET systematically presents mobility decrease with its MoS$_2$ layer thickness. It is also interesting to note that SS becomes accordingly larger with the thickness from 60 to 120 mV/dec. Gate leakage current $I_G$ was maintained to be ~1 pA and started to increase at 2.5 V gate voltage but could not be over 100 pA even at higher than 3 V. (Not shown here.) Besides above devices, we have again fabricated many other MoS$_2$-based MESFET devices on SiO$_2$/p$^+$-Si with different nanosheet thicknesses ranging from monolayer to 90 nm-thick bulk MoS$_2$, and confirmed aforementioned results (fig. S10 and Table S2). For all our MoS$_2$ MESFETs, it is also worth noting that a certain low threshold voltages exists since the MESFET gate opening is based on Schottky-driven operation. Such fixed threshold voltage should be marked as an important benefit of MESFET.

For theoretical understanding on the thickness-dependent mobility increase, we studied the respective electronic structures in bulk and monolayer MoS$_2$ using a first-principles density-functional method as implemented in the SIESTA code *(19)*. Figure 3, A and B show the band structures of bulk and monolayer MoS$_2$, respectively, drawn along the Γ-M-K-Γ line. These features of the bulk and the monolayer are consistent with previous theoretical reports *(3)*. As the mobility is proportional to the ratio of the carrier life time to effective mass $m^*$, we estimated the effective mass $m^* = \hbar^2(\partial^2\varepsilon\mu/\partial k^2)^{-1}$ at each conduction band minimum (CBM) from the conduction-band dispersions $\varepsilon(k)$, obtaining $m^* = 0.561\ m_e$ and $0.557\ m_e$ for bulk and monolayer $MoS_2$, respectively ($m_e$ is the electron mass in vacuum). Since the latter is only 1 % smaller than the former, we cannot attribute the increase in the mobility to any decrease in $m^*$. Thus, to elucidate the microscopic origin of the thickness dependence of the mobility, a scenario is still demanded for the carrier life time, which is a non-equilibrium property which requires information on scattering mechanism of carriers to defects, defect concentrations, etc. for a quantitative study. In our present work, we analyzed spatial shapes of the wavefunctions, $\psi_{n\vec{k}}(\vec{r})$, of the carriers at each CBM, to obtain an insight on any difference in scattering properties between monolayer and bulk. Figure 3, C and D show isosurface plots of the squared wavefunctions (electron probability density) at CBMs for bulk and monolayer $MoS_2$, respectively, taken at $|\psi_{n\vec{k}}(\vec{r})|^2 = 0.016\ \text{Å}^{-3}$. According to the results, sulfur $3p$ orbitals are more prominent in the CBM state of the bulk than of the monolayer. The butterfly-shaped sulfur $3p$ manifold in the CBM state of the bulk $MoS_2$ is the nearly equal mixture of $p_x$, $p_y$, and $p_z$ orbitals, while the $p_z$ component is missing at the CBM of the monolayer. In the CBM state of the bulk, the $4d$ manifold at Mo atoms consists of $d_{x^2-y^2}$, $d_{xy}$, and $d_{z^2}$ orbitals whose respective intensities decrease in that order, whereas the CBM state in the monolayer exhibits solely $d_{z^2}$ character at Mo atoms. Table 1 displays these features of the wavefunctions in another manner, showing the decompositions of wavefunctions at CBM. The atomic contributions, denoted by $C_{Mo}$ and $C_S$, are obtained from the Mulliken population analysis of each wavefunction *(20)*. A large $C_{Mo}$ or $C_S$ value implies a large electron occupation of Mo or S at CBM. According to Table 1, the compositional weight of the CBM wavefunction on S is significantly reduced from 45 % in the bulk to 16 % in the monolayer, which is approximately by three times. This suggests that if natural defects such as vacancies or some substitutional impurities (such as F, Cl, and Br) are located mostly at S sites in $MoS_2$ as unintentional donor, the electron carriers would much suffer a defect-induced scattering in the bulk $MoS_2$ but efficiently avoid such scattering in the monolayer; much higher electron mobility could thus be observed from the 1L $MoS_2$ (fig. S11 and Table S3). Moreover, when the electronic wavefunction at CBM reduces its weight at S sites and shrinks toward Mo sites, the conducting electrons in monolayer $MoS_2$ are slightly away from the $MoS_2/NiO_x$ interface and avoid any interface-involved scattering. (Fig. 3D)

Besides above-mentioned thickness effects on the mobility and carrier scattering reduction, we would mention that the $NiO_x$ gate electrode in direct contact with $MoS_2$ may maximize the gate-induced screening effect; the screening effects by high-k dielectric is known as a main origin of a high mobility in thin $MoS_2$ MISFETs but now we can obtain much higher screening effects by metallic $NiO_x$ electrode in MESFET and could demonstrate such high mobility as ~ 6000 $cm^2$/Vs. This value is even close to a theoretical mobility (6000 ~ 7000 $cm^2$/Vs at 100 K) in a free standing impurity-free $MoS_2$ which is limited by electron-phonon interaction *(21)*. The presence of high-k or metallic material in direct contact with $MoS_2$ has not been considered yet in the theory, but may produce significant screening effect of the electron-phonon interaction *(22)*. So, metallic $NiO_x$ electrode on thin $MoS_2$ leaves a room for a strong possibility of more enhanced mobility.

In summary, we have for the first time fabricated $MoS_2$-based MESFETs with $NiO_x$ Schottky electrode, where the carrier transport might not be interfered by the dielectric/$MoS_2$ channel interface of MISFETs. Whether fabricated on $SiO_2$/$p^+$-Si or glass substrate, our MESFETs demonstrate unprecedentedly high mobilities of 6000, 3500, and 2800 $cm^2$/Vs at low operational voltages with single-, double-, and triple-layered $MoS_2$, respectively. Theoretical analysis on the $MoS_2$ thickness-dependent mobility in MESFETs shows that weak and strong scatterings for the conducting CBM electrons respectively exist at S sites in 1L and bulk $MoS_2$. We conclude that $MoS_2$ nanosheets have intrinsically much higher potentials in their carrier transport speed performance than the previously reported values and $MoS_2$-based MESFETs are promising and useful for future nanoelectronics.

**Acknowledgments:**

The authors acknowledge the financial support from NRF (NRL program, No. 2014R1A2A1A010048), Nano-Material Technology Development Program (2012M3a7B4034985), NRF of Korea (Grant No. 2011-0018306), and KISTI supercomputing center (Project No. KSC-2013-C3-008).

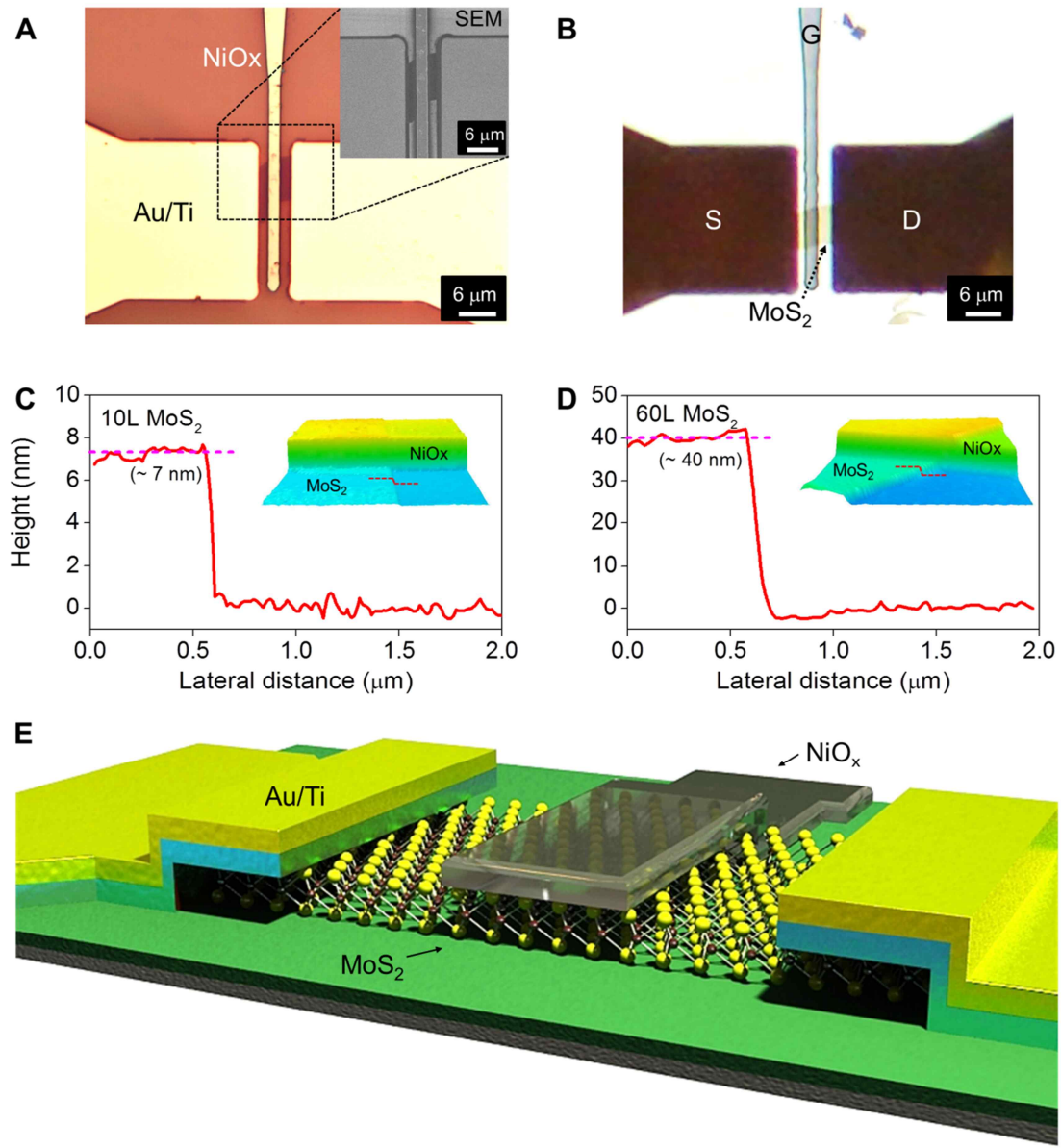

**Fig. 1**. MoS$_2$ thickness and MESFET structure on SiO$_2$/p$^+$-Si and glass substrate **(A)** OM and SEM images of 1L MoS$_2$ MESFET on SiO$_2$/p$^+$-Si. **(B)** OM image of MESFET on glass with 3L MoS$_2$, where we can clearly see MoS$_2$ through NiO$_x$ gate. L number was measured by Raman spectroscopy. **(C and D)** AFM topography results from 10L (~7 nm) and 60L (~40 nm) thick MoS$_2$ channel in MESFET **(E)** Schematic 3D view of MoS$_2$-based MESFET with 6 μm-long channel, 3 μm-long NiO$_x$ gate, and S/D region in contact with Ti.

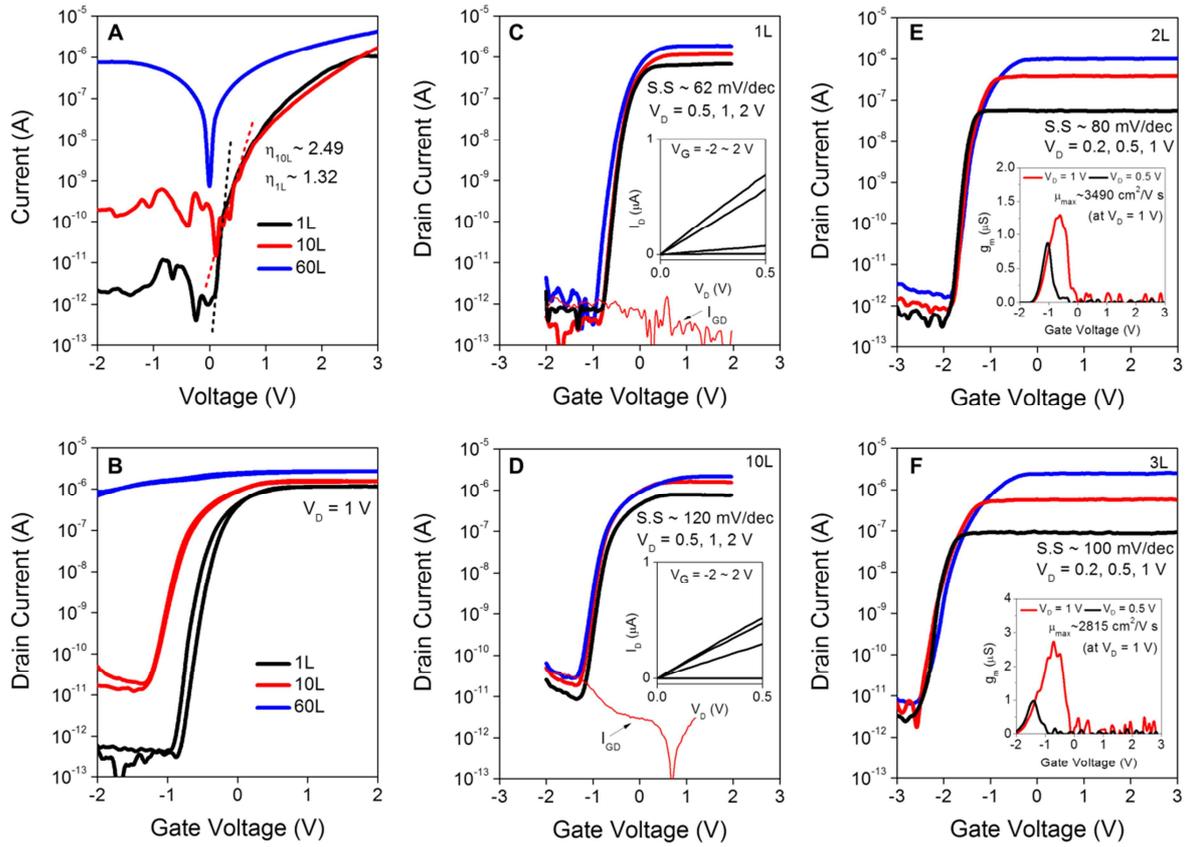

**Fig. 2.** Electron transport properties of our 1L, 10L, and 60L $MoS_2$ on $SiO_2/p^+$-Si **(A)** Schottky diode behavior was clearly observed from 1L and 10L $MoS_2$ with $NiO_x$ and Au/Ti electrodes. Since the Schottky behavior appears weaker with thicker $MoS_2$, the reverse leakage increases with the thickness. **(B)** Transfer curves of our MESFETs with 1L, 10L, and 60L $MoS_2$ obtained at a $V_D$ of 1 V; with the thickness, MESFET performance degrades. **(C)** Transfer curves of the MESFET with 1L $MoS_2$, as measured with $V_D$ increase; S.S. appears as good as ~ 62 mV/dec independent of $V_D$ and inset output curves confirms good ohmic contact. **(D)** Transfer curves of the MESFET with 10L $MoS_2$, as measured with $V_D$ increase; S.S. was ~120 mV/dec and inset shows the linear behavior of conducting electrons. **(E and F)** Transfer curves of the MESFETs with 2L and 3L $MoS_2$ as fabricated on glass; S.S. appears as good as 80 and 100 mV/dec, respectively and the inset curves show the transconductance plots to estimate the maximum field-effect mobility. According to the results, the mobility and SS performance gradually degrade with the thickness of $MoS_2$.

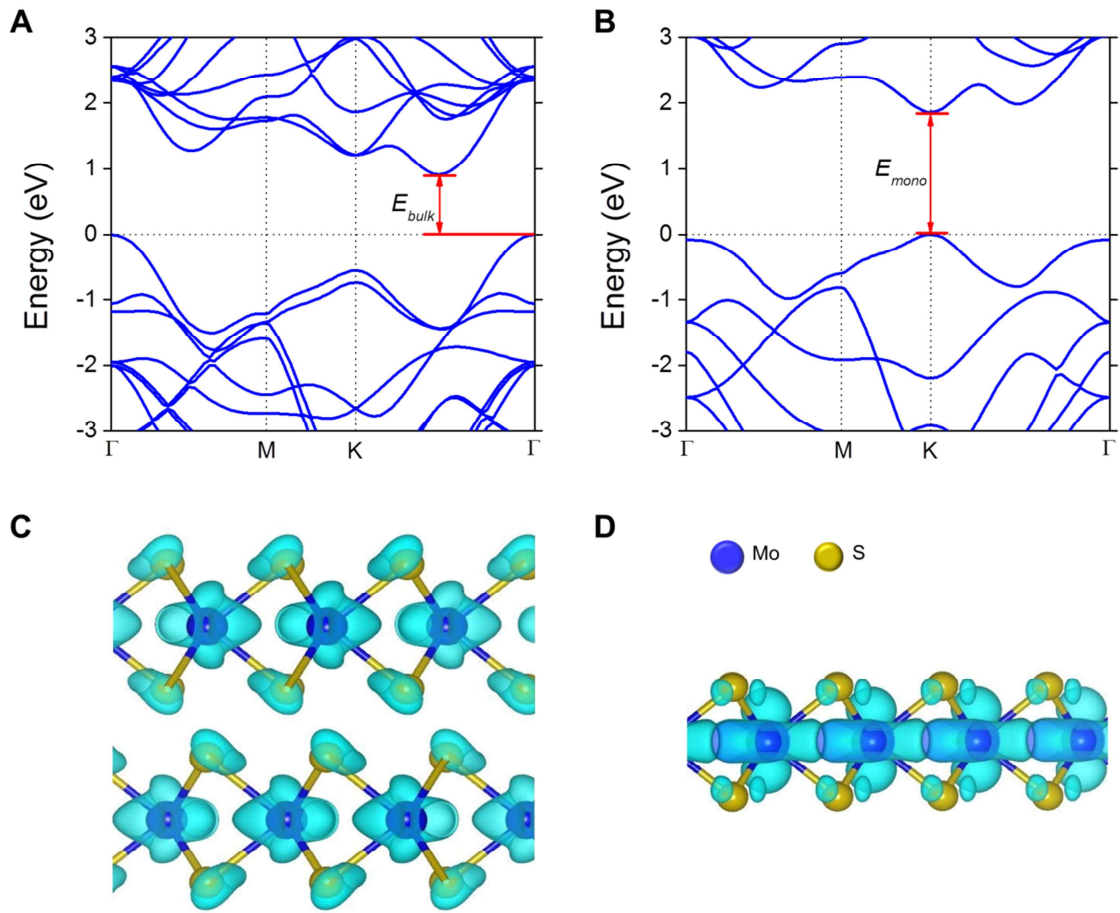

**Fig. 3**. Electronic structure of bulk and monolayer $MoS_2$. **(A)** Band structure of bulk $MoS_2$ along the high symmetry line. **(B)** Band structure of monolayer $MoS_2$ along the high symmetry line. **(C)** Isosurface plot of the squared wavefunction at CBM for bulk, taken at 0.016 Å$^{-3}$. **(D)** Isosurface plot of the squared wavefunction at CBM for monolayer, taken at 0.016 Å$^{-3}$. Blue and yellow spheres represent Mo and S atoms, respectively.

**Table 1.** Decomposition of wavefunctions at CBM in the bulk and the monolayer. The values of $C_{Mo}$ and $C_S$ are the contributions of Mo and S orbitals to CBM wavefunctions, respectively, as obtained from the Mulliken population analysis of each wavefunction.

| CBM | $C_{Mo}$ | $C_S$ | $m^*/m_e$ |
|---|---|---|---|
| Bulk | 0.55 | 0.45 | 0.561 |
| Monolayer | 0.84 | 0.16 | 0.557 |

## Supplementary Materials:

### Materials and Methods:

**Device fabrication.** 1-, 10-, and 60-L MoS$_2$ nanosheets were mechanically exfoliated from bulk MoS$_2$ crystals (SPI supplies, natural molybdenite) and transferred to 285 nm SiO$_2$ grown on heavily doped p-type silicon substrate (285 nm-thick SiO$_2$/p$^+$-Si) or to glass substrate by using a standard scotch tape method. For the source (S) and drain (D) ohmic contact electrodes Ti/Au (25/50 nm) were patterned (deposited at room temperature) on top of MoS$_2$ flakes by using photo-lithography, lift-off, and DC-sputter deposition processes. As the first layer, the Lift-Off-Layer (LOL: using a solution, LOL 2000, Micro Chemical) was coated, followed by thermal curing at 115 ºC for 2 min. Then photo-resist (PR : SPR 3612, Micro Chemical) as the second layer was coated and baked at 95 ºC for 2 min. The samples were exposed to UV light for 5 s under photo-mask aligner for S/D pattern. The samples were subsequently patterned with metal-ion-free (MIF) developer solution, and sequential DC-magnetron sputter-deposition of 30 nm-thin Ti and a 70-nm thin Au layers (Au/Ti) were carried out. For the lift-off process to finally define S/D electrode, acetone and LOL remover solvents were used. After that, the device was annealed at 250 ºC with N$_2$ flow in rapid thermal annealing (RTA) system, to remove polymer residue and simultaneously to reduce contact resistance. A 150 nm-thick NiO$_x$ Schottky contact gate electrode for MESFET (or Pt and Pd Schottky electrodes for diode) was patterned (deposited at room temperature) between the S/D electrodes by the same lithography and lift-off methods as performed for S/D contact, but for the deposition of NiO$_x$ we used the thermal evaporation of NiO$_x$ powder while Pt and Pd films were deposited by DC-magnetron sputtering. The channel length, L of our MESFET was 6 μm and the gate length of NiO$_x$, Pt, and Pd was 3 μm in channel direction.

**Measurements and Characterizations.** The thickness of MoS$_2$ flakes were firstly detected by optical microscope, then the exact layer number of each MoS$_2$ was characterized by AFM (XE-100, PSIA) and Raman spectroscopy. SEM was also necessary for the top view of our MESFET. All current-voltage (I–V) characteristics were measured by using a semiconductor parameter analyzer (HP 4155C, Agilent Technologies).

**Chemical potentials of pure elements for calculation of impurity formation energy.** Calculation of impurity formation energy requires chemical potential (i.e., reference energy) of pure elements. For chemical potentials of pure S, P, and As, we considered pure sulfur in orthorhombic structure (Space group No. 70, Fddd) with the lattice constants, $a$ = 10.4646Å, $b$ = 12.866 Å, and $c$ = 24.486 Å, pure phosphorus (Black P) in orthorhombic structure (Space group No. 64, Bmab) with the lattice constants, $a$ = 3.31Å, $b$ = 4.38Å, and $c$ = 10.5 Å, and pure arsenic in rhombohedral structure (Space group No. 166, R-3m) with the lattice constants, $a$ = $b$ = $c$ = 4.131 Å, and $α$ =54.1˚. For pure N, F, Cl, and Br, we considered diatomic-gas phases where the bond lengths of N$_2$, F$_2$, Cl$_2$ and Br$_2$ molecules are 1.45, 1.42, 1.99 and 2.28 Å, respectively.

## S1. Thickness characterization of a few layer MoS₂ by Raman spectroscopy

Raman spectroscopy from the monolayer shows the vibrational frequency peak difference of only 18.5 cm$^{-1}$ while the peak difference becomes wide but saturated with the layer thickness. In particular, the peak difference obtained from double layer MoS$_2$ is quite large to be 21.5 cm$^{-1}$ compared to that from the monolayer. *(1)*

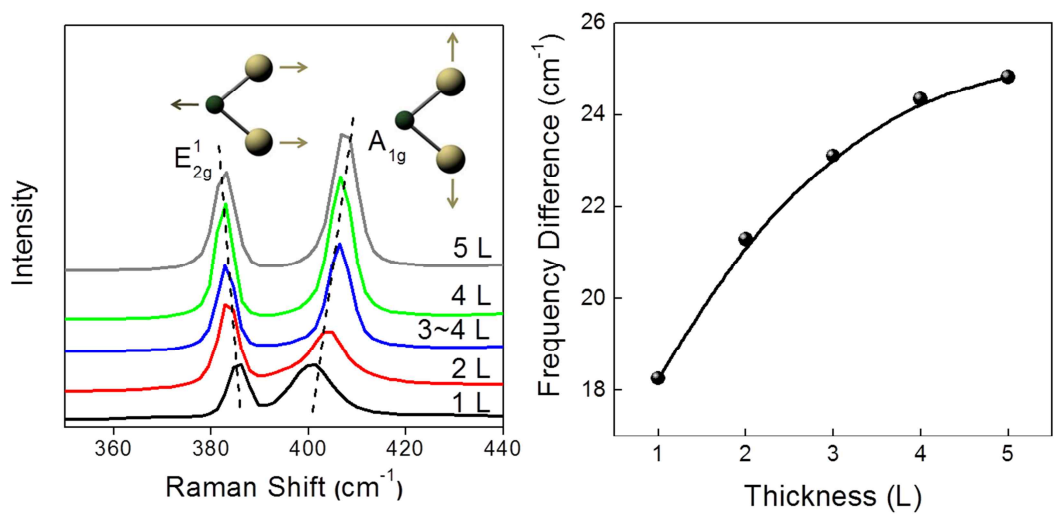

**Fig. S1.** Raman spectra obtained from mono- and a few layers of MoS$_2$, and resultant plot of frequency difference vs. layer number

## S2. Thick MoS₂-based MESFETs vs. monolayer thin MoS₂ MESFET

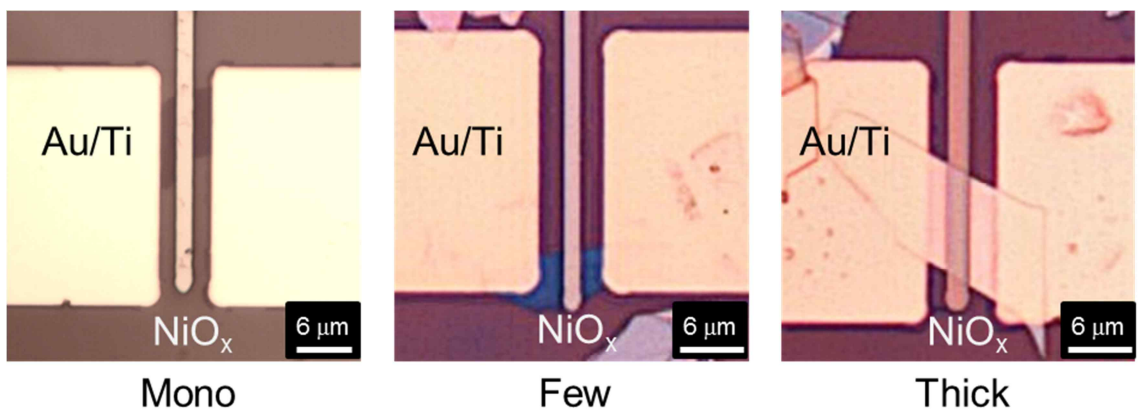

**Fig. S2.** Optical microscopies of 10- (left) and 60-L (right) thick MoS$_2$ nanosheets, which play as the channel of MESFET beneath NiO$_x$ gate; back ground substrate is 285 nm-thick SiO$_2$ on Si wafer. Interesting is that respective MoS$_2$ flakes have their own color contrast, and in particular the monolayer MoS$_2$ shows almost the same color contrast as the substrate while the 10 (few)-layer does blue. The 60L MoS$_2$ flake appears very thick, making a clear line contour on Au/Ti source/drain electrode. The present thick layer has a large width as well, compared to those of mono- and 10L MoS$_2$.

## S3. Device conditions for MESFET operation

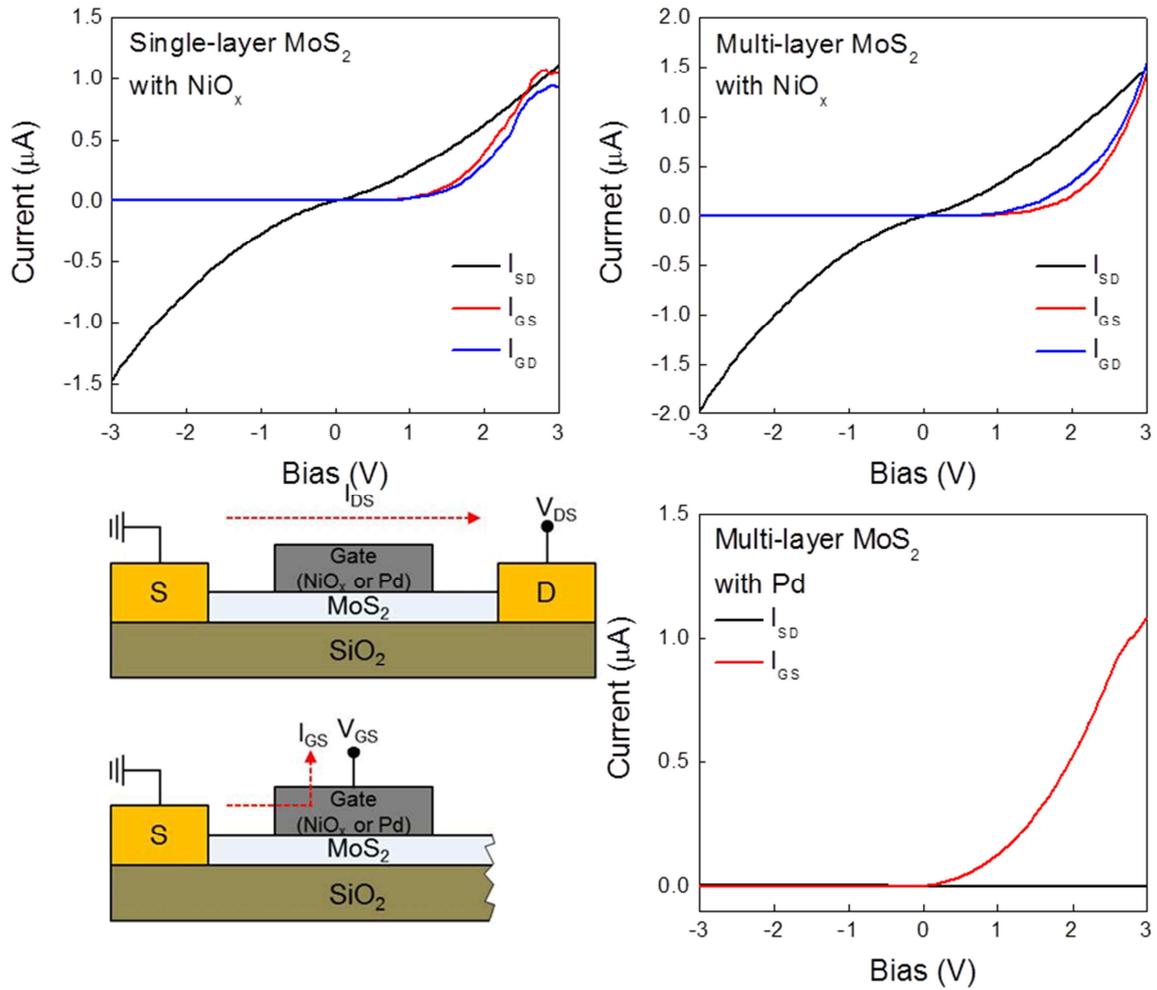

**Fig. S3.** I-V curves obtained from $NiO_x$/mono- and a few layers $MoS_2$ in device structures. As shown above, $I_{DS}$ current is always higher than $I_{GS}$ (or $I_{GD}$) Schottky current in a certain operation range satisfying the conditions for MESFET operation, which means that thermally-evaporated $NiO_x$ for gate could not completely deplete the channel $MoS_2$. In contrast, DC-sputter-deposited Pd, which contacts a few layer $MoS_2$, completely depletes the channel; $I_{DS}$ is always lower than $I_{GS}$ in the forward-bias voltage range. We suspect that the thermal-evaporation of $NiO_x$ powder might allow few angstrom-thin interlayer between $NiO_x$ and $MoS_2$, while $NiO_x$ has lower electron density than that of metal Pd.

## S4. Schottky effect expectation from NiO$_x$ electrode deposited on mono-, 10-, and bulk-like MoS$_2$

The work function of Ni-rich NiO$_x$ is expected to be 5.1 ~ 5.2 eV from the literature. *(2)* According to the diode-like I-V results of Fig. 2A, the 1L MoS$_2$ must have a larger Schottky barrier height than that of 10L while that of 60L MoS$_2$ should be minimal, in the interfacial contact with NiO$_x$. Below is a schematic energy diagram to be able to explain our results of Fig. 2A.

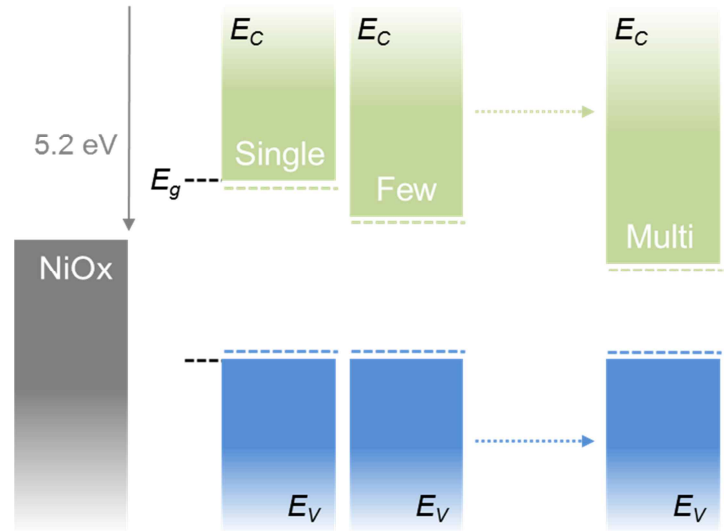

**Fig. S4.** Schematic energy band diagrams of monolayer and thick layer MoS$_2$ vs. the work function of NiO$_x$

Related to Fig. S4, I-V properties of the NiO$_x$/MoS$_2$ are determined by the well-known following equations with Schottky barrier height ($q\Phi_{Bn}$) and ideality factor ($\eta$).

$$I = I_0 \left( exp\left( \frac{q(V)}{\eta kT} \right) \right)$$

$$ln(I) = ln(I_0) + \frac{qV}{\eta kT} \quad (1)$$

$$I_0 = AA^* exp\left( \frac{-q\varphi_{Bn}}{kT} \right) \quad (2)$$

Where Richardson's constant for NiOx/MoS2, A* is unknown yet, k = 1.38 × 10-23 JK-1 and T = 297 K. The MoS2 thickness-dependent ideality factors were estimated by using the linear curve fitting of the log of diode current. Although Φ$_{Bn}$ has not been estimated in quantity here, it would decrease with the MoS2 thickness since the reverse-bias current Io appears increasing with the thickness according to Fig. 2A.

## S5. On/Off switching schemes in MoS$_2$ MESFETs

Below are the schematic illustrations on channel open (charge conduction between S and D: left figure, forward bias) and channel close (depletion: right, reverse bias). Although the MoS$_2$ nanosheet thickness is very thin in general, more rapid switching is expected from thinner sheet such as monolayer channel.

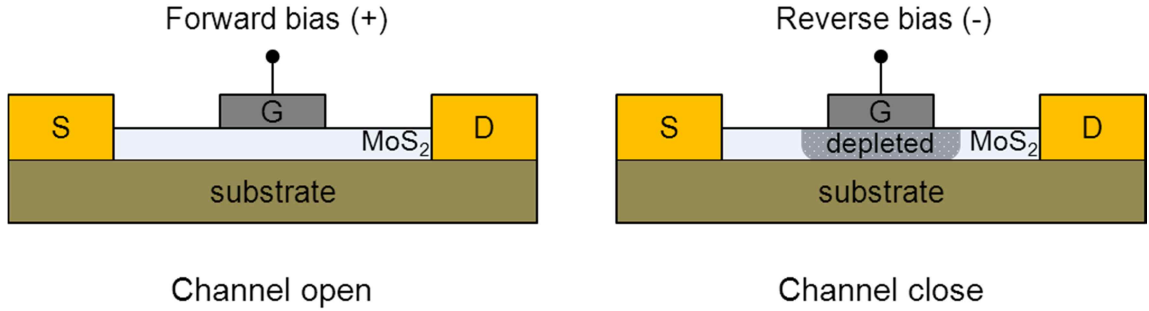

**Fig. S5.** Schematic illustrations for MESFET operations

## S6. V$_G$ dependent linear mobility in MoS$_2$ MESFET

As known in the literature *(3)*, the channel conductance $g_D$ for the linear region (at $V_D$ near 0 volt) is expressed with the transconductance $g_{Max}$ and $V_G$ as follows,

$$g_D = \frac{\partial I_D}{\partial V_D} = g_{Max}\left(1 - \left(\frac{V_i - V_G}{V_P}\right)^{\frac{1}{2}}\right)$$

, where $V_i$ is a built-in potential between metal and semiconductor, $V_p$ is minimum pinch-off voltage to deplete the MoS$_2$ thickness. Therefore, $g_{Max}$ is important here, to obtain a $V_G$-independent mobility when $V_G$ overcomes $V_i$ (on-stage).

Now, $g_{Max} = g_D$ as $V_G$ gets higher than $V_i$.

$$g_{Max} = \frac{\partial I_D}{\partial V_G} = \frac{qN_d\mu tW}{L}, \quad g_D = \frac{\partial I_D}{\partial V_D}$$

(at a small $V_D$ for linear region).

As a result, the mobility can be calculated as follows in a way,

$$\mu = \frac{L}{qN_d tW}(g_D) \cong \frac{L}{qN_d tW}\left(\frac{I_D}{V_D}\right)$$

,which follows the same concept as conductivity $\sigma = N_d q\mu$ in linear regime.

But, according to another way of calculation (for linear-to-saturation regime), we can calculate the channel mobilities using transconductance ($g_m$), since we have $g_m$ plots depending on $V_{DS}$. As known in the literature and paper, in the saturation regime, the channel mobility equation of MESFET can be expressed with $g_m$ as follows (3,4),

$$g_m = \frac{\partial I_D}{\partial V_G} = \frac{qN_d \mu t W}{L}$$

As a result, the channel mobility can be calculated as follows,

$$\mu = \frac{L g_{Max}}{q N_d t W}$$

Based on the I-V characteristics of 1L and 10L MoS$_2$ MESFET, we obtained voltage dependent $g_m$ plot.

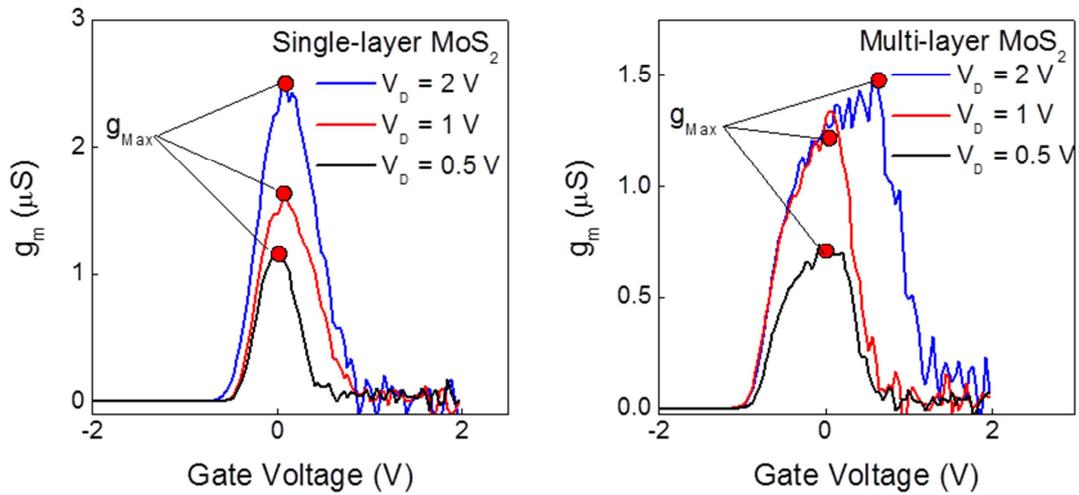

From above results, the maximum channel mobility, $\mu$ can be expressed with $g_{Max}$ as shown below. The $\mu$ of 1L becomes even higher than 6000 cm$^2$/Vs with $V_D$ increase while the mobility of 10L MESFET goes to ~ 600 cm$^2$/Vs at most.

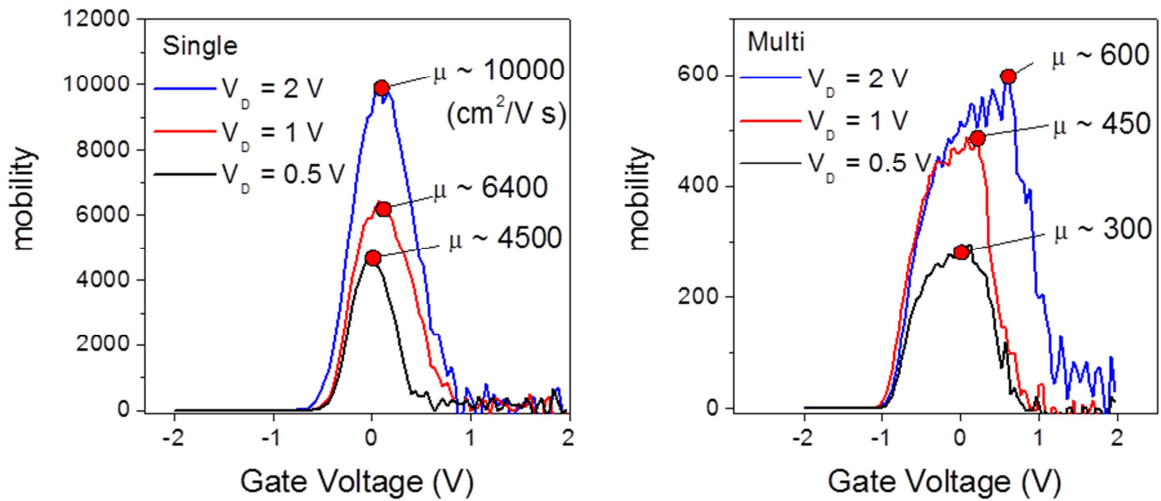

## S7. Padovani-Stratton parameter extraction from the reverse-bias I-V equation

We estimated the mobility values of our MoS$_2$ sheets (10L) by using the Padovani-Stratton parameter that should be extracted from the following equations related to the reverse-biased Schottky barrier (5,6), and we also estimated the electron density, $n$ value

$$I = SJ = SJ_s \exp\left(V\left(\frac{q}{kT} - \frac{1}{E_0}\right)\right) \quad (1)$$

where $J$ is the current density through the Schottky barrier, $S$ is the contact area associated with this barrier, $V$ is the bias voltage, $q$ is the magnitude of electronic charge, $k$ is the Boltzmann constant, $T$ is the absolute temperature, $J_s$ is a slowly varying function of applied bias under a reverse bias condition,

and

$$E_0 = E_{00} \coth\left(\frac{qE_{00}}{kT}\right) \quad (2)$$

with

$$E_{00} = \frac{qh}{4\pi}\left(\frac{N_d}{m^* \varepsilon_r \varepsilon_o}\right) \quad (3)$$

where $E_{00}$ is the Padovani–Stratton parameter, which is significant in tunneling theory, $N_d$ is the donor density at the metal/semiconductor interface, $m^*$ and $\varepsilon_r$ are the effective mass and relative permittivity of the semiconductor, respectively, $h$ is Planck's constant, and $\varepsilon_o$ is the vacuum permittivity. The logarithmic plot of the current as a function of the bias voltage gives a slope corresponding with $[q/kT - 1/E_0]$. The electron concentration, $n$ (or $N_d$) can be calculated though $E_0$ and $E_{00}$. Below are the linear and logarithmic plots and table for 10L MoS$_2$ Schottky diode.

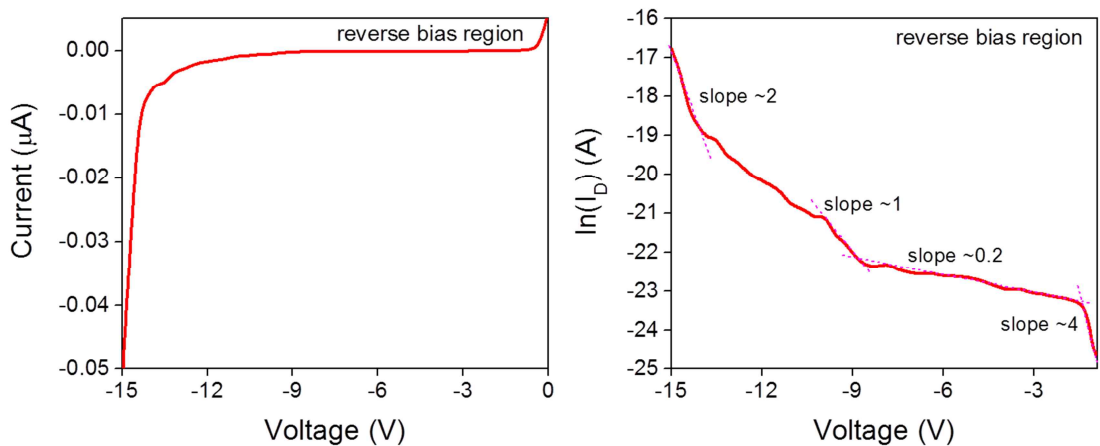

**Fig. S7.** The plots of $I_D$ vs. reverse voltage and logarithmic ($I_D$) vs. reverse voltage

**Table S1.** Slope, $q/kT-1/E_0$ extracted from the logarithmic $I_D$ plot; the electron concentration, $n$ (or $N_d$) can be calculated though $E_0$ and $E_{00}$.

| Slope | $E_o$  | $E_{oo}$ | $E_{oo}^2$ | $N_d$   |
|-------|--------|----------|------------|---------|
| 0.2   | 0.0262 | 0.00395  | 1.56e-5    | 2.35e15 |
| 0.5   | 0.0264 | 0.00559  | 3.12e-5    | 4.71e15 |
| 1.0   | 0.0268 | 0.00792  | 6.27e-5    | 9.45e15 |
| 1.5   | 0.0272 | 0.00972  | 9.45e-5    | 1.42e16 |
| 2.0   | 0.0274 | 0.01051  | 1.10e-4    | 1.66e16 |
| 2.5   | 0.0278 | 0.01193  | 1.42e-4    | 2.15e16 |
| 3.0   | 0.0282 | 0.01321  | 1.75e-4    | 2.63e16 |
| 3.5   | 0.0286 | 0.01438  | 2.07e-4    | 3.12e16 |
| 4.0   | 0.0290 | 0.01547  | 2.40e-4    | 3.70e16 |

Since our measurement was focused in the range between 0 and -3 V, we consider the slope from 0.2 to 4.0, so $N_d = 1.42 \times 10^{16}$ cm$^{-3}$ would be an average, $n$ of 10L MoS$_2$.

## S8. Reflection terahertz time-domain spectroscopy (RTHz-TDS) and results for electron density in undoped 60 μm-thick bulk $MoS_2$

Terahertz time-domain spectroscopy (THz-TDS) *(7)* has been recently introduced for condensed matter physics for semiconductors *(8)*, superconductors *(9)*, and 2-dimensional electron gas materials *(10, 11)*. The carrier dynamics of semiconductors is sensitive to the carrier concentration, the effective mass, and the plasma frequency in the terahertz range (0.1 – 10 THz). The Drude model predicts that the A.C. conductivity in the terahertz range will exhibit a Lorentzian conductivity spectrum (the real part), which is directly related to the carrier density. Here, we present on our reflection terahertz time-domain spectroscopy (RTHz-TDS) of bulk $MoS_2$ based on the reflection phase error correction technique with the maximum entropy method (MEM). Fig. S8A is the experimental setup of our RTHz-TDS (Teraview TPS 3000). For the reflection setup, the emitter and the receiver parts of femto second laser beams were guided through two optical fiber cables into the two separate photoconductive emitter and receiver placed near the bulk $MoS_2$ (about 5 mm × 5 mm × 0.06 mm). The distance between the photoconductive emitter (or receiver) and bulk $MoS_2$ was about 50mm and the angle of incidence was about 15˚.

Fig. S8B presents the reflected terahertz time-domain electric field of bulk $MoS_2$ (red) and evaporated Au reference mirror (black) (evaporated *in situ* on the same bulk $MoS_2$). Despite employment of evaporated gold thin film reference mirror on bulk $MoS_2$, the MEM is indispensible to phase error correction due to the inevitable relative displacement between the bulk $MoS_2$ and Au reference mirror. Following the standard MEM procedure *(12-15)*, we started with the experimental phase $\theta_{exp}(\omega)$, as a function of angular frequency $\omega$ and calculated the MEM phase $\theta_{MEM}(\omega)$ from the experimentally measured terahertz reflectivity $R(\omega)$. The difference between $\theta_{exp}(\omega)$ and $\theta_{MEM}(\omega)$ is assumed to be linear in angular frequency and the corresponding linear slope $\alpha$ is determined by differentiating $\theta_{exp}(\omega) - \theta_{MEM}(\omega)$ with respect to angular frequency $\omega$, which in turn provides the MEM corrected phase $\theta_{cor}(\omega) = \theta_{exp}(\omega) - \alpha\omega$.

Fig. S8c shows the optical conductivity of bulk $MoS_2$ (black) and three fitting curves in the terahertz range. The optical conductivity of bulk $MoS_2$ is well fitted (red) to its Drude (blue) and the Lorentz (green) components. The D.C. limit conductivity about 2 $\Omega^{-1}cm^{-1}$ from the Drude fit is consistent with the measured D.C. conductivity of about 1.5 $\Omega^{-1}cm^{-1}$ (open diamond). From the Drude model, we obtained the plasma frequency $\omega_p$ = 41 $cm^{-1}$ and scattering rate $\gamma$ = 8.0 $cm^{-1}$ and calculate the carrier concentration of $MoS_2$, $n = 1 \times 10^{16}$ $cm^{-3}$. Following this procedure, we calculate the carrier concentration $n$ defined by

$$n = \frac{\omega_p^2 m^* \varepsilon_0}{e^2}$$

where $m^*$ is the effective mass, $\varepsilon_0$ the electric permittivity of free space, and $e$ is the electronic charge. Here, the effective mass is assumed to be ~ 0.6 $m_e$ (electron mass $m_e$) *(16)*. The Lorentzian resonance peak about 56.4 $cm^{-1}$ appears at the position of inactive $B_{2g}^2$ phonon mode incidentally *(17)*. We conjecture the activation of an optical phonon mode due to symmetry breaking, possibly due to interlayer defects and stacking layer mismatch.

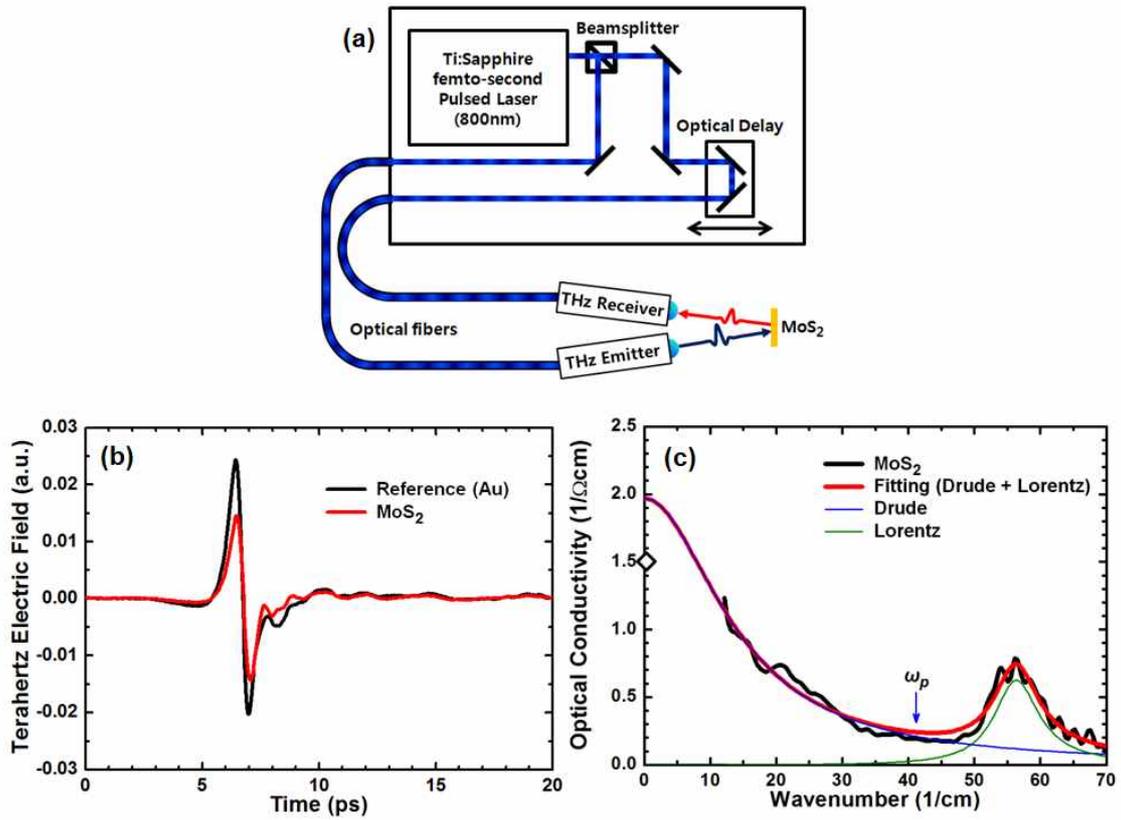

**Fig. S8.** (A) The reflection setup of terahertz time-domain spectroscopy. (B) The reflected terahertz electric field of Au reference mirror (black) and bulk $MoS_2$ (red). (C) The real part of optical conductivity of $MoS_2$ bulk (black) and fitting curves. The Drude + Lorentz fitting curve (red) indicates a combination of Drude part (blue) and Lorentz (green). The open diamond is the measured DC conductivity for a $60LMoS_2$ sample.

According to above terahertz spectroscopy results, the expected D.C. conductivity is ~ 2 $\Omega^{-1}cm^{-1}$ from the Drude fit, which is similar to ON-state conduction 10L $MoS_2$ MESFET (1.5 $\Omega^{-1}cm^{-1}$) while 60L $MoS_2$ MESFET shows 0.8 $\Omega^{-1}cm^{-1}$ by manual calculation. Since Padovani-Stratton method does bring $1.42 \times 10^{16}$ $cm^{-3}$ as the electron density while the terahertz technique does $1.1 \times 10^{16}$ $cm^{-3}$, we believe our electron density-based linear mobility value is quite reasonable.

## S9. 4-probe Hall measurement for 20nm-thick MoS$_2$ flake on glass

We have fabricated a few samples on glass for conventional Hall measurement through 4 times of sequential lift-off processes. Contact metal was Au/Ti. Shown below are AFM thickness measurement results and 4 probe Hall sample with the 20 nm thickness. As Hall measurement results, we achieved important values of Carrier Concentration ($N$) = 2.2 × 10$^{16}$ (cm$^{-3}$), $N·t$ (thickness) ~ 5.0 × 10$^{10}$ (cm$^{-2}$), and electron Hall mobility ($\mu$) ~ 130 cm$^2$/Vs. We also show the MESFET device on glass (the zoom-out version of Fig. 1B in the main manuscript) below for information.

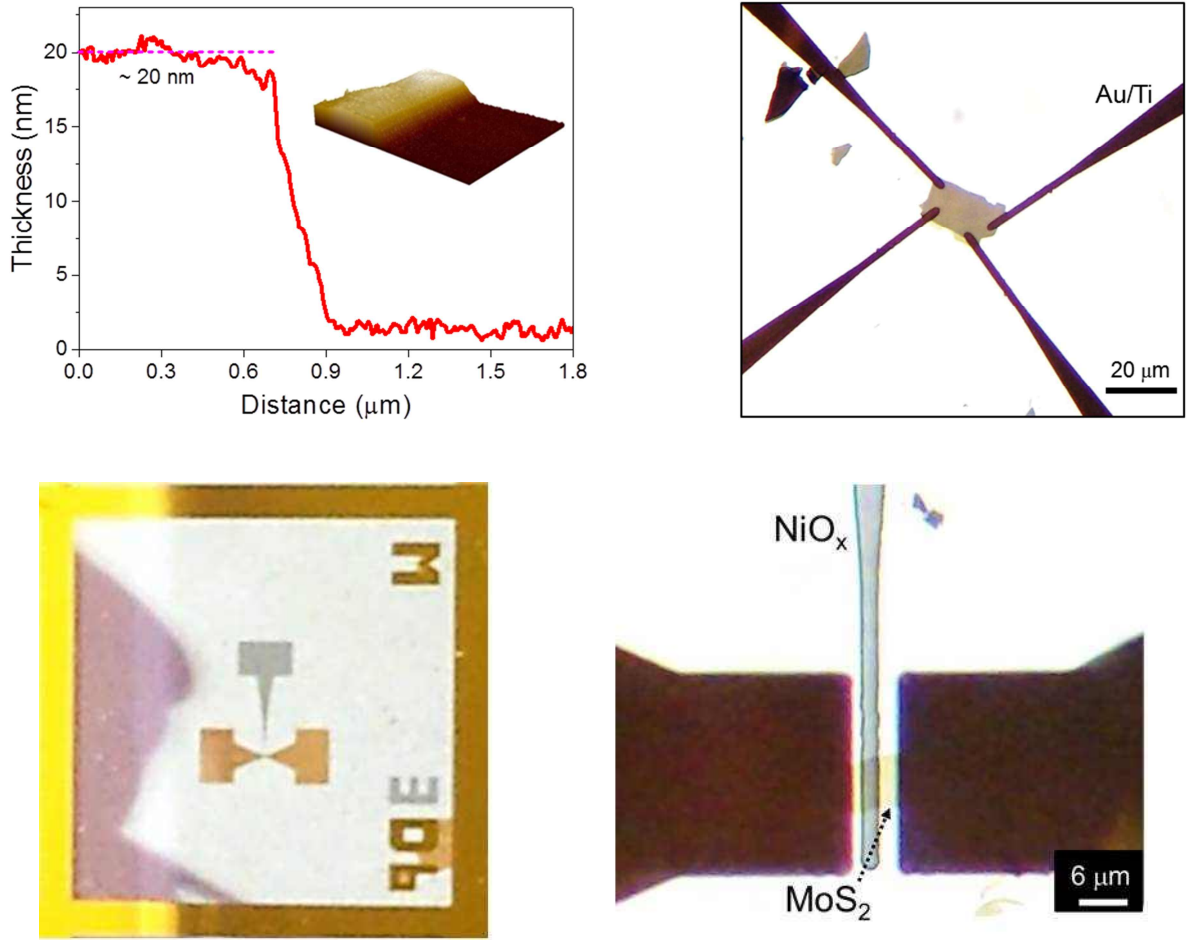

**Fig. S9.** AFM thickness profile of 20 nm-thick MoS$_2$ and its 4 probe Hall measurement. The sample on glass was as large as 10~20 μm in size. The down-side photos shows our MESFET on glass substrate, where we can see MoS$_2$ through NiO$_x$ gate.

## S10. Example transfer characteristics of our MESFETs with thin and thick MoS$_2$ channel layers

We attempted to fabricate many other MoS$_2$-based MESFET devices with NiO$_x$ gate and confirmed that thinner MoS$_2$ causes lower OFF current and S.S as well, which is shown below from the transfer curves and summary table. MESFETS with thick MoS$_2$ layer mostly display high $I_D$ current but also high OFF current. The thickness-enhanced OFF current signifies that the electron affinity and Fermi energy of n-type MoS$_2$ more approach to the work function of Ti contact as its thickness gets larger. As a result, the Schottky effects gradually disappear with the gradual MoS$_2$ thickness increase, while ON current increases due to the increased channel thickness.

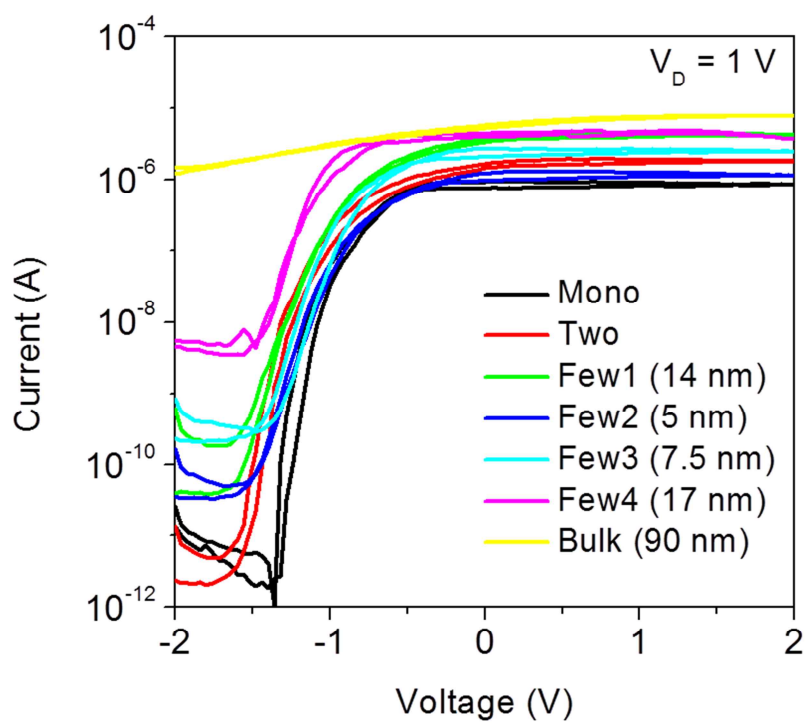

**Fig. S10.** (A) Transfer curves of MESFETs with several MoS$_2$ thicknesses

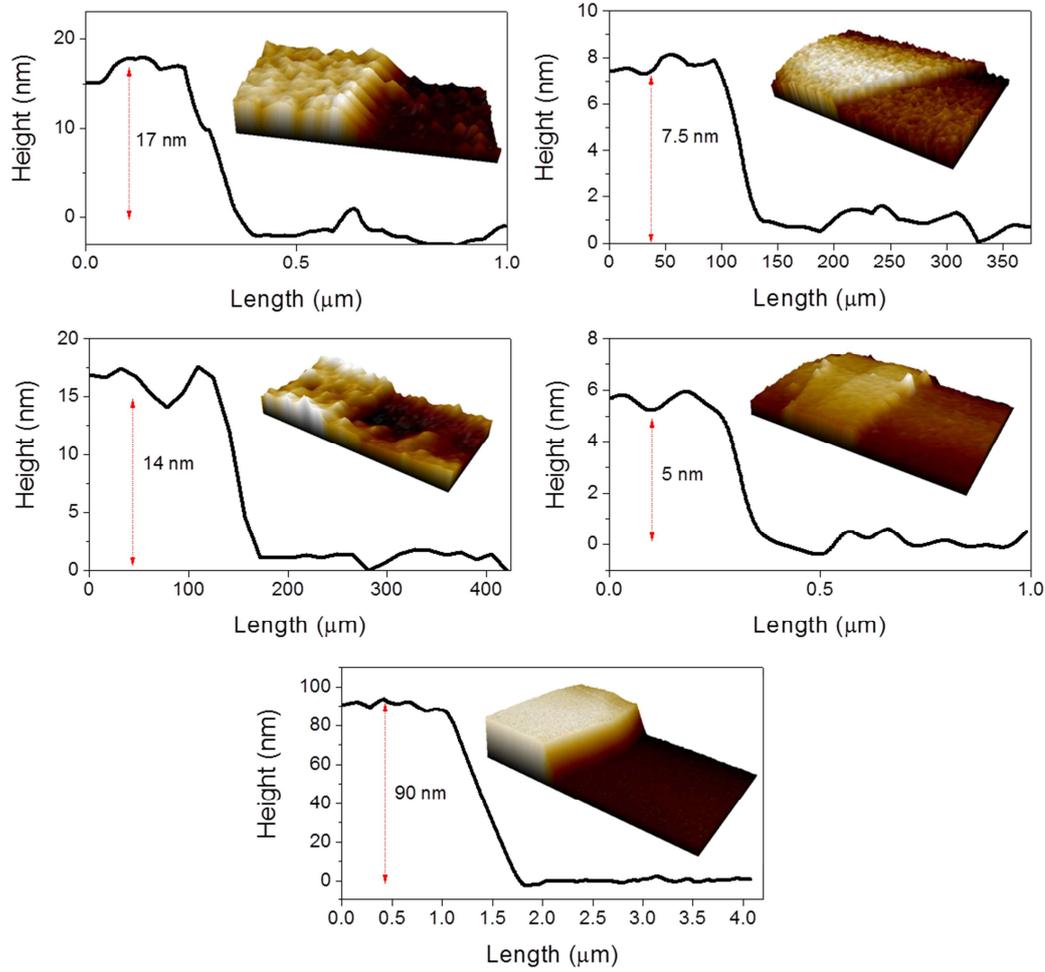

**Fig. S10.** (B) AFM-profiled MoS$_2$ thickness for the MESFETs presented in Fig. S8B

**Table S2.** Summarized properties of the MESFETS presented in Fig. S8A

| # of layer | $I_{ON}$(A) | $I_{OFF}$(A) | ON/ OFF ratio | S.S (mV/dec) | μ (cm$^2$/Vs) |
|---|---|---|---|---|---|
| Mono | $0.9\times10^{-6}$ | $1.4\times10^{-12}$ | ~$10^6$ | 60 | 6912 |
| Two | $1.7\times10^{-6}$ | $3.0\times10^{-12}$ | ~$7\times10^5$ | 80 | 3663 |
| Few1 (14 nm) | $4.5\times10^{-6}$ | $4.0\times10^{-11}$ | ~$10^5$ | 120 | 912 |
| Few2 (5 nm) | $1.4\times10^{-6}$ | $4.5\times10^{-11}$ | ~$3\times10^4$ | 120 | 869 |
| Few3 (7.5 nm) | $2.7\times10^{-6}$ | $2.1\times10^{-10}$ | ~$10^4$ | 150 | 1495 |
| Few4 (17 nm) | $4.8\times10^{-6}$ | $4.0\times10^{-9}$ | ~$10^3$ | 140 | 1500 |
| Bulk (90 nm) | $7.8\times10^{-6}$ | $1.0\times10^{-6}$ | ~10 | x | 321(=$\mu_{linear}$) |

One thing noteworthy from above is that 5~17 nm-thin MoS$_2$ shows similar μ values

## S11. Unintentional electron donor sources

Previous scanning tunneling microscopy experiments reported that MoS$_2$ has S vacancies as native defects *(18, 19)*, and the electron irradiation can create the sulfur vacancies *(20)*. We thus performed density-functional calculations of bulk Mo$_{18}$S$_{35}$ and monolayer Mo$_{16}$S$_{31}$ expecting sulfur vacancy as an electron source, but we only obtaineda result that both bulk and monolayer MoS$_2$ become p-type semiconductors with S vacancies. Therefore, excluding the S vacancies from the donor candidates, we turned to choose S-substituting halogen impurities (e.g. F, Cl, and Br) as the most feasible electron source of natural MoS$_2$ in our analysis. Table S3 shows our calculated formation energies of substitutional impurities $X$ ($X$ = halogen and pnictogen elements) for S in monolayer MoS$_2$. According to our results, the formation of F impurity appears spontaneous with negative formation energy and that of Cl impurity requires smaller formation energy than others. We thus regarded F and Cl impurities as good promising candidates for electron donor in MoS$_2$ and also as the scattering source to explain the mobility difference in the two MESFET systems of monolayer and 60 layer (bulk) MoS$_2$; the difference originates from different scattering rate (or life time) of CBM electrons provided by impurities at S sites.

**Table S3.** Formation energy ($E^F$) of substitutional impurity $X$ for S ($X$ = halogens and pnictogens) in monolayer MoS$_2$. The formation energy $E^F$ per impurity atomis calculated by using the formula, $E^F = E[\text{Mo}_9\text{S}_{17}X] - E[\text{Mo}_9\text{S}_{18}] - \mu[X] + \mu[S]$, where $E[\text{Mo}_9\text{S}_{17}X]$ is the total energy of the system with a substitutional impurity $X$ in a 3×3×1 supercell of monolayer MoS$_2$, $E[\text{Mo}_9\text{S}_{18}]$ is the total energy of a 3×3×1 supercell of perfect monolayer MoS$_2$, and $\mu[X]$ and $\mu[S]$ are the chemical potentials of impurity $X$ and sulfur, respectively (See Method for more details).

| impurity (n-type) | F | Cl | Br |
|---|---|---|---|
| $E^F$ (eV) | -0.58 | 1.50 | 2.51 |
| impurity (p-type) | N | P | As |
| $E^F$ (eV) | 1.88 | 1.98 | 2.50 |